\documentclass[11pt,a4paper]{article}

\usepackage[nohead, nomarginpar, margin=1in, foot=.25in]{geometry}

\usepackage{graphicx}

\usepackage{csquotes}

\usepackage{amsmath}

\usepackage[T1]{fontenc}
\usepackage[utf8]{inputenc}
\usepackage{authblk}

\usepackage[LGRgreek]{mathastext}

\usepackage{cite}
\date{}

\def\@listcomma@comma{\@ifnum{\@tempcnta>\tw@}{,}{}}

\begin{document}
\makeatletter
  \def\title@font{\Large\bfseries}
  \let\ltx@maketitle\@maketitle
  \def\@maketitle{\bgroup%
    \let\ltx@title\@title%
    \def\@title{\resizebox{\textwidth}{!}{%
      \mbox{\title@font\ltx@title}%
    }}%
    \ltx@maketitle%
  \egroup}
\makeatother

\title{Production of D-mesons in $p$+$p$ and $p$+Pb collisions at LHC energies}
\author[1,\footnote{ramachandrabaral@yahoo.co.in}] {R.C.Baral}
\author[1,2] {S.K.Tripathy}
\author[1] {M.Younus}
\author[2] {Z. Naik}
\author[1] {P.K.Sahu}

\affil[1] {Institute of Physics, Bhubaneswar-751005, India}
\affil[2] {Sambalpur University, Burla-768019, India}

\maketitle

\begin{abstract}
We present theoretical model comparison with published ALICE results for
D-mesons (D$^0$, D$^+$ and D$^{*+}$) in $p$+$p$ collisions at $\sqrt{s}$ = 7 TeV and $p$+Pb
collisions at $\sqrt{s_{NN}}$ = 5.02 TeV.
Event generator HIJING, transport calculation of 
AMPT and calculations from NLO(MNR) and FONLL have been used for this study.
We found that HIJING and AMPT model predictions are matching with published D-meson cross-sections in $p$+$p$ collisions, while both under-predict the same in  $p$+Pb collisions. Attempts were made to explain the $R_{pPb}$ data
using NLO-pQCD(MNR), FONLL and other above mentioned models.
\end{abstract}


\section{Introduction}
Relativistic heavy ion collisions at the RHIC \cite{rhic} and the LHC \cite{lhc} have given 
rise to a new phase of matter. When two heavy
ions collide, a system of de-confined gluons and quarks within a very small volume is created. 
The initial energy density within this volume is found to be much larger than nuclear 
ground state energy density.  
This state of matter as we know today is called Quark Gluon Plasma (QGP) \cite{qgp1,qgp2}.  The
study of QGP is particularly important as it aims to produce a condition, which resembles the period when
universe was only a few microseconds old.  However, since this exotic system 
created in the experiments exists only for a very short period of time and is not directly observable, only signals
originating from the matter itself that survive and are measured after the collisions can provide
a window into the nature of the QGP \cite{qgprev1,qgprev2}. With high statistics data already 
accumulated at the Large Hadron Collider at CERN, the scientific community has an enormous
task to analyse, and explain these observations and
extract information about the properties of the QGP. These
analyses are also leading the way for additional measurements and 
will become available for studies with all the major experiments, 
like STAR \cite{star, ref_2}, ALICE \cite{alice, ref_3} and proposed CBM at FAIR \cite{cbm, ref_4}.

One of the prominent signatures coming out of the QGP phase is jet quenching \cite{ref_1}. 
High momentum hadron spectra are observed to be highly
suppressed relative to those in $p$+$p$ collisions \cite{suppress1,suppress2}, 
suggesting a quenching effect due to deconfined matter. A similar effect is observed for 
high $p_T$ charm or beauty quarks with most recent results showing suppression of D or
B mesons to same order as that of light partons \cite{exp1, sd1}. However before going into 
hot and dense matter effects, it is absolute necessary to fix the baseline for such 
observations. In heavy ion scenario, $p$+$p$ collisions serve as the baseline for such observations, 
assuming that no nuclear effects are present when $p$+$p$ is scaled to $p$+Pb or Pb+Pb data only by a factor. 
On the other hand, it has been suggested that the modification in spectra of the observed 
particles in the heavy ion collision have effects of cold nuclear matter \cite{cnm} before formation of QGP which 
are often masked by hot and dense matter effects. So it is important to discern the contributions 
of the cold nuclear effects from all other effects due to QGP on the final particle spectra. $p$+Pb 
collisions give us a unique opportunity to study these initial nuclear effects. The so called effect 
due to shadowing 
has been playing a role in the particle production scenario for a very long time. 
With the assumption that any nucleus is not just any conglomeration of protons is the very 
essence of this phenomenon. With 
LHC achieving its top collider energy, it may not be possible  to overlook the shadowing features 
affecting the high gluon density within the nucleus.
This phenomenon is also represented mathematically as shadowing ratio, $R_s \equiv F_A(x,Q^2)/(A*F_p(x,Q^2))$, 
and has been found to deviate from unity as explained in early literatures \cite{shadow1}, which makes this 
phenomenon as one of the most prominent feature 
of cold nuclear effects. On the other hand, another phenomenon that may affect the final particle spectra is 
multiple re-scattering of the colliding nucleons or their partons. This effect is known as Cronin 
effect \cite{cronin1}. This particular feature had been observed in the RHIC energy for non-photonic electrons' 
nuclear modification data, which shows an enhancement in the charm spectrum below $p_T < 4.0 $ GeV \cite{phenix1}. 
The results suggest that this particular effect may be observed 
in the low and mid-p$_T$ regions and may not be much effective in higher side of the momentum. We will come back to these two points later in our work.

Now let us move over to heavy quarks. A heavy quark owing to its large mass is produced much 
before the formation of quark gluon plasma \cite{charm1}. It is also 
believed that heavy quarks remain free to 
probe thermalized medium without carrying any prior effects due to nucleus. 
From the recent result of $p$+Pb data and earlier $d$+Au data \cite{charm2} on particle production, 
the value of $R_{pPb}$ deviates from unity by almost 
15$\%$ mostly in low and mid-p$_T$ regions, which shows a considerable cold nuclear matter effect on heavy 
quark production \cite{charm3}. The current work aims to highlight some of 
these initial nuclear effects on measured heavy meson spectra.

This paper is organised as follows. In the section \ref{Models_used} we discuss the various models  
employed for studying D-meson cross-section
 in $p$+$p$ and $p$+Pb collisions at  $\sqrt{s}$  = 7 TeV and $\sqrt{s_{NN}}$ = 5.02 TeV, respectively. 
In the section \ref{Results_and_discussion} we discuss our results with these models. 
Then we summarise our work in section \ref{Summary}.


\section{Models used}
\label{Models_used}

\subsection{The HIJING model}
\label{Hijing}
HIJING (Heavy-Ion Jet INteraction Generator)\cite{hij} is a Monte Carlo  model designed mainly to explore the range of possible initial conditions that may occur in nuclear collisions at collider energies and to produce output that can be compared directly with a wide variety of nuclear collider experimental observables. 
The main features included in HIJING are as follows. \\
The formulation of HIJING is guided by Lund FRITIOF \cite{Lund} and Dual Parton Model \cite{dpm} for soft nuclear reaction at 
intermediate energy ($\sqrt{s_{NN}} \leq 20$ GeV). 
Multiple low $p_{T}$ exchanges among the end point constituents are included to model initial state interactions. 
The PYTHIA \cite{pythia} guides the pQCD processes where multiple minijet production with initial and final state radiation are involved. To reproduce $p$+A or A+A results, the Eikonal formalism is used to calculate the number of minijets per inelastic $p$+$p$ collision. 
The model uses three-parameter Woods-Saxon nuclear density determined by electron scattering data \cite{paper45}. 
A diffuse nuclear geometry decides the impact parameter dependence of the number of binary collisions \cite{paper6}. 

The cross section for charm production formalism at the leading order is written as \cite{paper22}
\begin{equation}
\frac{d\sigma_{cc}^{pp}}  {dp_{T}^{2} dy_1 dy_2} = K \sum_{a,b} x_1 f_a(x_1,p_T^2) \ x_2 f_b(x_2,p_T^2)
\times \frac{d\hat{\sigma}_{ab}} {d\hat{t}} , \\
\end{equation}
here $a,b$ are the parton species, $y_1, y_2$ are the rapidities of the scattered partons, and $x_1, x_2$ are the fraction of momentum carried by the initial partons. A factor $K$, of value 2.0 has been used to account roughly for the higher order corrections. In HIJING, the parton structure functions, $f_a(x_1,p_T^2)$ are the  Duke-Owens  \cite{paper21} structure function set 1 and this is also implemented in PYTHIA.
For the nuclear effect in A+A and $p$+A collisions, model follows the {\it A} dependence of the shadowing proposed in Ref. \citenum{paper48,paper50} and uses its parameterization as
\begin{equation} 
 \label{eqn:shadow}
\begin{split}
R_{A}(x)  \equiv \frac{f_{a/A}(x)}  {A \ f_{a/N}(x)}  = \ & 1+1.19 \ ln^{1/6} A [ x^3 - 1.5(x_0+x_L)x^2+3x_0x_Lx]   \\
 & - \Big[\alpha_A(r) - \frac{1.08(A^{1/3} -1) } {ln(A+1)} \sqrt{x} \Big] e^{-x^{2}/x_0^2} \ , \\ 
\end{split}
\end{equation}
 and $ \alpha_A(r) = 0.1 ( A^{1/3} -1 )\frac{4}{3} \sqrt{1-r^2/R_A^2}. $ \\ 
Here r is the transverse distance of the interacting nucleon from its nucleus centre and $R_A$ is the radius of the nucleus, and $x_0=0.1$ and $x_L =0.7$. The most important nuclear dependence term is proportional to $\alpha_A(r) $ in Eq.\ref{eqn:shadow}, which determines the shadowing for $x<x_0$, and the rest gives the overall very slow {\it A} dependence nuclear effect on the structure function for $x>x_L$.

We have used HIJING version 1.41.

\subsection{The AMPT model}
\label{Ampt}

\noindent A Multiphase Transport Model (AMPT) \cite{ampt_model} is a hybrid transport model, which was developed to address non-equilibrium many body dynamics. Initially it was designed to describe physics of $p$+A and A+A collisions for centre of mass energy from 5 GeV to 5.5 TeV. Outline of this model are as follows.

Initial distribution of nucleons inside a nucleus is taken from HIJING and is Woods-Saxon in nature. Scattering among them are treated with Eikonal formalism. If momentum transfer ($Q^{2}$) is greater than a cut off momentum ($p_{0}$), then these processes produce minijet partons and treated with PYTHIA model. Reverse ($Q^{2} < p_{0} $) leads to production of strings. Depending on spin and flavor of excited strings, they get converted into partons without any further interaction. If those strings or partons satisfy minimum distance conditions ($\leq \sqrt{\sigma/\pi}, \ \sigma$ being cross section for partonic two-body scattering), then they undergo interactions that are dealt by Zhang's Parton Cascade (ZPC) model \cite{zpc}. Once these partons stop interacting, nearest two partons form a meson or that of three form a baryon using a quark coalescence model. Cascade of resultant hadrons is dealt by a relativistic transport model, ART \cite{art_1, art_2}, which includes baryon-baryon, baryon-meson and meson-meson elastic and inelastic scatterings.

This version of AMPT, known as string melting has been used for the current study (version 26t5). There is another version referred as default AMPT model, where instead of quark coalescence, string fragmentation method is adopted.

\subsection{The NLO model}
\label{Nlo}

The next-to-leading order, NLO-pQCD(MNR)\cite{MNR} model used in the 
present work has been successfully used before to 
produce \textit{c$\bar{\textit c}$} pair 
cross-sections in $p$+$p$ collisions at most of the available collider energies \cite{jamil}. 
Consequently the model can be used to produce various heavy quark spectra and can be utilised 
further to study various hot and dense nuclear matter effects (as in Pb+Pb and Au+Au collisions) 
and cold nuclear matter effects (as in $p$+Pb and $d$+Au collisions). In the present work, 
we have used the calculations to produce D-meson spectra 
for $p$+$p$ collisions at $\sqrt{s}$ = 7 TeV in order to check the consistencies of our calculations. 
In the next step, the calculations have been repeated for $p$+Pb at $\sqrt{s_{NN}}$ = 5.02 TeV including shadowing 
effects as one of the initial cold nuclear effects \cite{shadow2,ramona1}. Let us now move to a 
brief description of the calculations:

The $p_T$ differential spectrum of heavy quarks produced in $p$+$p$ collisions is defined in general 
as \cite{jamil,younus}
\begin{equation}
{E_1}{E_2}\frac{d\sigma}{d^{3}p_1 d^{3}p_2} = \frac{d\sigma}{dy_1 dy_2 d^{2}p_{T_1} d^{2}p_{T_2}}\  ,
\label{ini11}
\end{equation}

where $y_1$ and $y_2$ are the rapidities of heavy quark and anti-quark 
and {\bf\it p}$_{ \textit {\bf T}_i}$ are their transverse momenta.\\
In the above
\begin{equation}
\begin{split}
 \frac{d\sigma}{dy_1 dy_2 d^{2}p_{T_1} d^{2}p_{T_2}} = & \ 2 x_{a}x_{b}\sum_{ij} \bigg[ f^{(a)}_{i}(x_{a},Q^{2})f_{j}^{(b)}(x_{b},Q^{2}) \frac {d\hat{\sigma}_{ij}(\hat{s},\hat{t},\hat{u})} {d\hat{t}} \\
& +  f_{j}^{(a)}(x_{a},Q^{2})f_{i}^{(b)}(x_{b},Q^{2}) \frac{d\hat{\sigma}_{ij}(\hat{s},\hat{u},\hat{t})}{d\hat{t}} \bigg] /(1+\delta_{ij}) \ ,\\
\label{ini2}
\end{split}
\end{equation}

where $x_{a} $ and $x_{b} $ are the fractions of the momenta carried by the partons 
from their interacting parent hadrons.

We have used CTEQ6.6 structure function \cite{cteq66} as obtained using LHAPDF library for $p$+$p$ system and 
added EPS09 \cite{eps09} shadowing parameterization, 
to incorporate the initial nuclear effects on the parton densities for $p$+Pb system.

 The differential cross-section for partonic interactions, $d\hat{\sigma}_{ij}/d\hat{t}$ 
is given by
\begin{equation}
\frac{d\hat{\sigma}_{ij}(\hat{s},\hat{t},\hat{u})}{d\hat{t}} = \frac{\left|M\right|^{2}}{16\pi\hat{s}^{2}} ,
\label{dsdt}
\end{equation}
where $\left|M\right|^{2}$ (See Ref. \citenum{combridge}) is the invariant amplitude for various 
partonic sub-processes both for leading order (LO) and next-to-leading order (NLO) processes as follows: \\
The physical sub-processes included for the leading order, 
$\cal{O}$ $(\alpha_{s}^{2}) $ production of heavy quarks are
\begin{equation}
\begin{split}
 g+g & \rightarrow Q+\overline{Q} \ and  \\
 q+\bar{q} & \rightarrow Q+\overline{Q} \ . 
\end{split}
\end{equation}
At next-to-leading order, $\cal{O}$ $(\alpha_{s}^{3})$ subprocesses 
included are as follows
\begin{equation}
\begin{split}
g+g & \rightarrow Q+\overline{Q}+g \ , \\
q+\bar{q} & \rightarrow Q+\overline{Q}+g \ and \\
g+q(\bar{q}) & \rightarrow Q+\overline{Q}+q(\bar{q}) .
\end{split}
\end{equation}

Next we discuss re-scattering processes within the nucleus.
A parton may also undergo multiple hard scattering or a 
 nucleon instead undergo multiple soft re-scattering within the cold nucleus in cases of $p$+A or A+A collisions. 
This is commonly referred as Cronin effects \cite{cronin1,accardi1}. 
These re-scatterings may lead to momentum broadening of the 
interacting partons and change the final heavy quark spectrum. 
This would also give rise to deviations of $R_{pPb}$ from unity and is considered as another 
form of cold nuclear matter effect. We feel that its contribution apart from shadowing to the heavy meson spectra, 
when compared to $p$+$p$ collisions, can be discerned with the precise state-of-the-art experiments 
designed at LHC-CERN and RHIC-BNL. However, it was earlier suggested that this effect may vanish at large 
transverse momentum region or high collider energies \cite{accardi2,sharma1,levai1}, but 
may be visible in the low and mid $p_T$ region and is slowly emerging as 
a subject of contemporary interests in heavy ion collisions. 
The details of our implementations of the calculations are taken from 
Ref. \citenum{accardi1,gyulassy1}. 

We can now discuss briefly about one of the mechanisms used from the above references. 
Starting with parton density functions, which can be defined as
\begin{equation}
f^{(a)}_{i}(x_{a},Q^{2},k_T^2)=f_{i}^{(a)}(x_{a},Q^{2}).g_{p/A}(k_T^2) \ , \\
\end{equation}

where $g_{p/A}(k_T^2)\propto exp[-k_T^2/\pi\,.\langle k_T^2\rangle_{pp/pA}]$ and $\langle k_T^2\rangle_{pA}= \langle k_T^2\rangle_{pp} + \langle k_T^2\rangle_A$ .\\

The effective transverse momentum kick, 
$\langle k_T^2\rangle_{pA}$, following leads from Ref. ~\citenum{accardi2} and \citenum{gyulassy1},
 is obtained by adding $\langle k_T^2\rangle_A$ as a consequence of 
series of re-scattering, to the intrinsic $\langle k_T^2\rangle_{pp}$. Our preliminary assumption of 
taking this summation however doesn't extrapolate $p+A$ system exactly to $p+p$ scenario. We are currently 
looking to improve upon this assumption. The $\langle k_T^2\rangle_A$ can be assumed as 
\begin{equation}
 \langle k_T^2\rangle_A = \delta^{2} . n.\ln \bigg( 1+\frac{p_T^2}{\delta^2/c} \bigg)
\end{equation}
where the parameters $\delta^2/c$, average squared momentum kick per scattering 
and $n=L_A/\lambda\,,L_A=4R_A/3$, average number of re-scattering, are used from Ref. \citenum{sharma1,gyulassy1}.

With the implementation of the above features, we can next fragment the charm
momentum both from $p$+A and $p$+$p$
collisions into D-mesons, as
D-mesons data are readily verifiable from experiments. Schematically,
this can be shown as
\begin{equation}
E\frac{d^3\sigma}{d^3p}=E_Q\frac{d^3\sigma}{d^3p_Q}\otimes D(Q\rightarrow H_M) ,
\label{eq18}
\end{equation}
where the fragmentation of the heavy quark $Q$ into the heavy-meson $H_M$ is
described by the function $D_D(z)$. We have assumed that distribution of
$D(z)$, w.r.t. $z$, where $z=p_D/p_c$, is used to calculate total $D$-mesons and
is given by
\begin{equation}
D^{(c)}_D(z)=\frac{n_D}{z[1-1/z-\epsilon_p/(1-z)]^2} ,
\label{eq19}
\end{equation}
where $\epsilon_p$ is the Peterson parameter $\simeq \ $0.12 and
is taken from Ref. \citenum{peterson1}.
The normalization condition satisfied by the fragmentation
function is
\begin{equation}
\int_0^1 \, dz \, D(z)=1 .
\label{eq20}
\end{equation}

\subsection{The FONLL model}
\label{Fonll}

As mentioned in the literatures, FONLL \cite{cacciari1} has been used to calculate D-mesons spectra 
for LHC energies and earlier estimations have shown 
that FONLL calculation is able to explain various heavy quark observables particularly 
transverse momentum spectra of heavy mesons with remarkable accuracies.
The p$_T$ spectra of heavy quarks produced in $p$+$p$ collisions as in Eq. \ref{ini11} can be written as

\begin{equation}
E_c\frac{d\sigma}{d^{3}p_c dy_{c}} =
\int d^{3}p_{\bar{c}} dy_{\bar{c}} \frac{d\sigma^{pp\rightarrow c\bar{c}}} {d^{3}p_{c} d^{3}p_{\bar{c}} dy_{c} dy_{\bar{c}}} \ ,
\label{ini1}
\end{equation}
where $y_c$ and $y_{\bar{c}}$ are the rapidities of heavy 
quark and anti-quark and {\bf\it p}$_{ \textit {\bf T}_i}$ are their transverse momenta.

The above distribution is evaluated at the
Fixed-Order Next-to-Leading-Logarithmic (FONLL) level,
implemented in Ref \citenum{cacciari1}. In addition to the
full fixed-order NLO result, the FONLL calculation also resums large perturbative terms proportional to $\alpha^n_s=\log^k(p_T/m_c)$
at all orders with next-to-leading-logarithmic (NLL) accuracy, where
$m_c$(= 1.5 GeV) is the heavy quark mass. 
Here too, we have used CTEQ6.6 parton structure function and EPS09 
shadowing parametrization for our calculations. 

The charm fragmentation function developed by Cacciari \textit{et al.} \cite{cacciari2} is used in the present work. This
depends on the parameter $r$ (See Ref. \citenum{nason1}) with
the values of the parameters defined in the above references and fitted with $e\mathord{+} e\mathord{-}$ 
spectra data. 
Bottom
fragmentation instead depends on the parameter $\alpha_B$ in a
functional form given by Kartvelishvili \textit{et al}.\cite{kartvelishvili1} It is worth noting that
using the Peterson \textit{et al}. fragmentation function,
gives a different result than that of
fragmentation in FONLL \cite{cacciari1}.

\section{Results and discussion}
\label{Results_and_discussion}
\noindent ALICE has recently published results on D-meson in 
$p$+$p$ \cite{alice_pp_7} and $p$+Pb \cite{alice_ppb_5020} collisions. 
Keeping on view of that, events are generated at $\sqrt{s}$ = 7 TeV and 
$\sqrt{s_{NN}}$ = 5.02 TeV using all the above models, i.e. HIJING, AMPT, 
NLO and FONLL. For $p$+$p$ system, the study is based on the mid rapidity 
region, i.e., $|y_{cms}| <$ 0.5, where as for $p$+Pb system it is in the 
rapidity range -0.96 $<$ y$_{cms}$ $<$ 0.04 . We have ensured that no D-meson is 
coming from B-meson.

Normalised $p$+$p$ yield was divided by $T_{pp} = 1.39 \times 10^{-5}$ $\mu b^{-1}$ 
to obtain cross-section, while that for $p$+Pb $T_{pPb}$ is $9.8334 \times 10^{-5}$ $\mu b^{-1}$ 
(calculated in Ref. \citenum{yellow1}).

From calculated cross-section, Nuclear modification factor $(R_{pPb})$ can be defined as follows:
\begin{equation}
\scalebox{1.2} {$R_{pPb}$ = $\frac{ (\frac{d\sigma}{dp_{T}})_{pPb} }{A\times(\frac{d\sigma}{dp_{T}})_{pp}}$},
\end{equation}
where {\it A} is the mass number of a nucleus (e.g., for Pb it is 208). Here we have used $p$+$p$ collisions as baseline at $\sqrt{s_{NN}}$= 5.02 TeV. We will discuss these various cold nuclear 
matter effects on our results in the following sections.

\vspace{0.18cm}
\begin{figure}[!ht]
\centering
\includegraphics[scale=0.41]{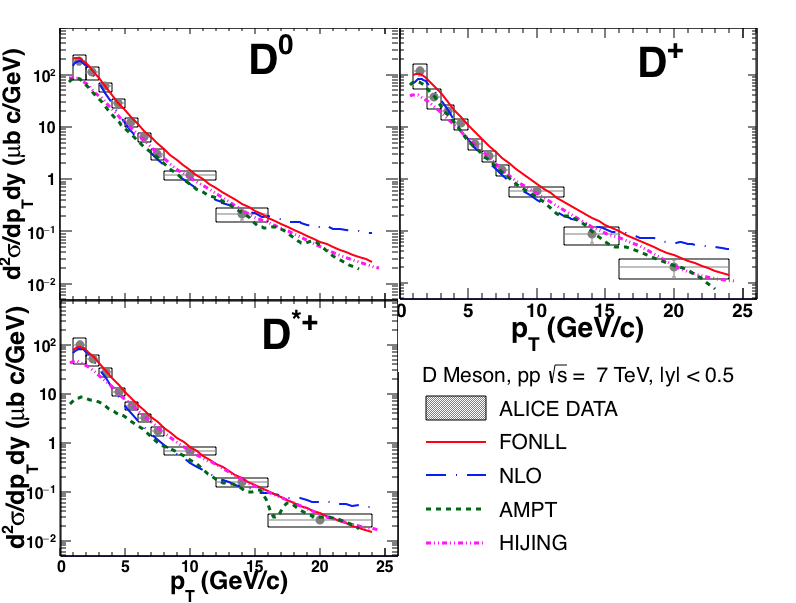}
\caption{\small(Color online) $p_{T}$ differential inclusive 
production cross-section of D-meson in $p$+$p$ $\sqrt{s}$ = 7 TeV. 
Solid markers represent the ALICE data points \cite{alice_pp_7}. 
Statistical errors are in bars while systematic errors are in boxes. 
Small dash-dotted line (Magenta), dashed line (Green), long dash-dotted line 
(Blue) and solid line (Red) represent HIJING, AMPT, NLO and FONLL results, respectively.}
\label{fig:ampt_pp_7}
\end{figure}

Figure \ref{fig:ampt_pp_7} shows transverse momentum (p$_T$) 
differential production cross-section of D$^0$, D$^+$ and D$^{*+}$ 
mesons in $p$+$p$ collisions at $\sqrt{s}$ = 7 TeV. Except for few low $p_{T}$ bin, 
HIJING explains data within the uncertainties. Similar trend is followed by AMPT 
for $D^{0}$ and $D^{+}$, but it poorly explains cross section of $D^{*+} $ for $p_{T} < 10 $ GeV/c. 
Apart from the direct production of $D^{0}$ and $D^{+}$, we have incorporated 
contributions from other resonance decays. However, there is no decay contribution of 
other particles for the production of $D^{*+}$. 
From figure \ref{fig:ampt_pp_7}, we may say that both String Fragmentation and quark 
coalescence based simulation models (HIJING and AMPT respectively) are able to explain 
results from $p$+$p$ collision data.
In addition to that, there might be some additional production mechanism is needed for AMPT especially at low $p_{T}$, which might add up to the $D^{*+}$ cross-section. NLO 
results explains the data within error bars up to $p_{T} < 15 $ GeV/c, 
but over estimates the results at higher $p_T$ region. This may be due to 
the large NLO contributions adopted in the model formalism. Its shape is 
different from other simulations, which might be due to its dependence on 
renormalisation and fragmentation scale factors. FONLL at its next-to-leading 
calculations explains data very well for all $p_{T}$ region.

\vspace{0.18cm}
\begin{figure}[!ht]
\centering
\includegraphics[scale=0.41]{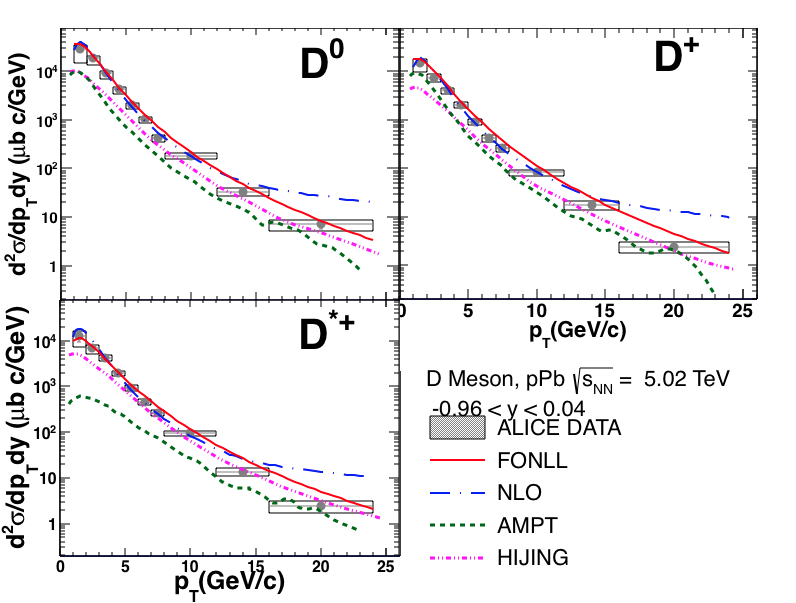}
\caption{\small(Color online) $p_{T}$ differential inclusive 
production cross-section of D-meson in $p$+Pb data at $\sqrt{s_{NN}}$ = 5.02 TeV. 
Solid markers represent the ALICE data points\cite{alice_ppb_5020}. 
Statistical errors are in bars while systematic errors are in boxes. 
Small dash-dotted line (Magenta), dashed line (Green), 
long dash-dotted line (Blue) and solid line (Red) 
represent HIJING, AMPT, NLO and FONLL results, respectively.}
\label{fig:ampt_pbpb_5020}
\end{figure}

Figure \ref{fig:ampt_pbpb_5020} is same as that of Figure \ref{fig:ampt_pp_7}, 
but for $p$+Pb system at $\sqrt{s_{NN}}$= 5.02 TeV. Here HIJING under-predicts the data 
for $p_{T} < 7 $ GeV/c. So we may think that cold nuclear shadowing effect of Pb as 
implemented in this model might have suppressed the yield to a large extent. 
AMPT under-predicts the data for all $p_{T}$ region for $D^{0}$ and $D^{+}$, 
but have same miss-match as that of $p$+$p$ for the case of $D^{*0}$. On contrary to HIJING, 
AMPT shows a smaller production cross-section for $D^{0}$ and $D^{+}$ in $p$+Pb system 
in its mechanism irrespective of nuclear shadowing effect. NLO in $p$+Pb likewise over 
estimates the cross-section for $p_{T} >15 $ GeV/c. FONLL explains data for all $p_{T}$ 
to very good extent.

\vspace{0.18cm}
\begin{figure}[!ht]
\centering
\includegraphics[scale=0.41]{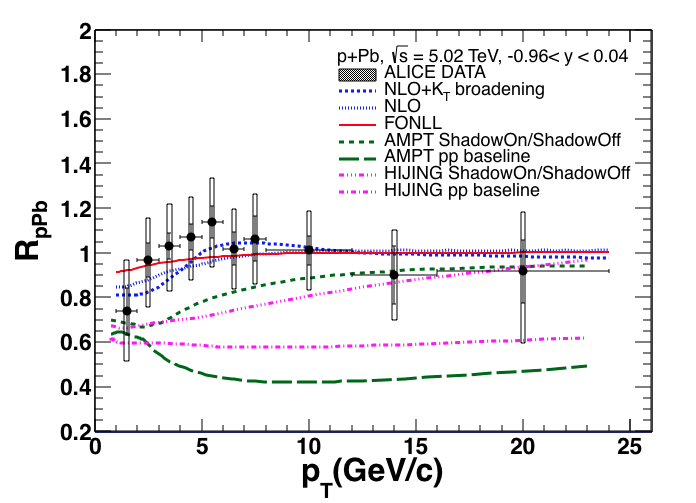}
\caption{\small(Color online) Nuclear modification factor for D-meson in $p$+Pb at $\sqrt{s_{NN}}$= 5.02 TeV. 
Solid markers represent the ALICE data points \cite{alice_ppb_5020}. 
Statistical errors are in bars while systematic errors are in boxes.  Dash-dotted line (Magenta), 
dashed line (Green), solid line (Red) and dotted line (Blue) represent HIJING, AMPT, FONLL and NLO results respectively. AMPT ShadowOn/ShadowOff or HIJING ShadowOn/ShadowOff represent results from taking nuclear effect shadowing on in numerator to off in denominator (other nuclear effects kept un-changed) in the same system , i.e. $p$+Pb at $\sqrt{s_{NN}}$= 5.02 TeV.}
\label{fig:ampt_Rppb_5020}
\end{figure}

Figure \ref{fig:ampt_Rppb_5020} shows p$_T$-dependence of average $R_{pPb}$ of D$^0$, D$^+$ and 
D$^{*+}$ mesons in $p$+Pb data at $\sqrt{s_{NN}}$ =  5.02  TeV. The calculations 
from HIJING and AMPT are showing prominent cold nuclear matter (CNM) effects such as 
shadowing, EMC \cite{emc}, and multi-parton scattering effects, for the entire $p_{T}$ range. 
The results under-estimate the magnitude and trend of experimental data. The reasons behind such 
large CNM effects implemented in these calculations are being investigated and will be reported 
in our future publications. Besides having quark coalescence as parton production mechanisms in AMPT than that of string fragmentation in HIJING, AMPT has additional partonic and hadronic transport parts which 
have both elastic and inelastic scatterings. This may also be the reason that $R_{pPb}$ from AMPT is lower than that of HIJING. Next, in case of NLO, which has its nuclear shadowing feature, and in addition, 
it has momentum broadening effect (Cronin) 
due to re-scattering. Both the results with and without the momentum broadening 
effects are shown in the plot. The corresponding result with additional momentum 
broadening are closer to the trend of the data within 
its error bars. The result using NLO without broadening is closer to 
unity with suppression at the low $p_T$ region due to shadowing and shows 
a difference in the shape of the curve from the one including the broadening effect. 
We may recall that a similar enhancement in trend of $R_{dAu}$ for $\pi_0$ meson 
has also been reported for $d$+Au collisions at $\sqrt{s_{NN}}$ = 5.5 TeV by 
M. Gyulassy \textit{et al}. (see Ref. \citenum{gyulassy1}). FONLL with shadowing features only too gives very 
small shadowing effect for $p_{T} <$ 10 GeV/c and remains close to unity.

Using AMPT and HIJING, to show the effects of shadowing exclusively on nuclear 
modification factor and also difference between $p+p$ and $p$+Pb (shadow-off) as baselines, we further calculated 
$R_{pPb}$ as following:\\
\begin{equation}
\scalebox{1.2} {$R_{pPb}$ = $\frac{ (\frac{d\sigma}{dp_{T}})_{pPb}^{ShadowOn}}{(\frac{d\sigma}{dp_{T}})_{pPb}^{ShadowOff}}$} \\
\end{equation}

Here we have turned on shadowing effect in numerator and turned it off in denominator (while other nuclear effects 
like multi-parton scattering etc. are present in both) in the same system, i.e. $p$+Pb at $\sqrt{s_{NN}}$= 5.02 TeV.  
As we can see from Figure \ref{fig:ampt_Rppb_5020} that taking $p$+Pb (shadow-off) as the baseline we see considerable nuclear effects such as shadowing  particularly at the low and intermediate 
$p_T$ regions, while any other effects due to Pb nucleus is cancelled 
both from numerator and denominator of the ratio. The results however differ much from calculations using 
$p+p$ baseline (AMPT and HIJING), suggesting greater effects of multi-parton scattering than shadowing 
etc. on the final D meson spectra. 
\section{Summary} 
\label{Summary}
\noindent We have carried out D-meson study in simulation models like HIJING and AMPT and 
compared our results with published ALICE data for $p$+$p$ collisions at $\sqrt{s}$ = 7 TeV
and $p$+Pb collisions at $\sqrt{s_{NN}}$ = 5.02 TeV.
We have also compared with the results from next-to-leading order calculations from FONLL and NLO.

Irrespective of shadowing effect included in both the models, AMPT shows lower value of $R_{pPb}$ 
compared to HIJING above $p_{T} = 2.5 $ GeV/c. So we may conclude that 
magnitude of $R_{pPb}$ in AMPT due to its additional partonic and hadronic transport parts differs 
from the same in HIJING. And for resonance particle D$^{*+}$, 
additional mechanism is needed in AMPT to explain its production cross-section. 
More details in this direction will be reported in our future study.

Since $R_{pPb}$ in all our calculations deviates from unity, thus there is initial cold 
nuclear matter effect playing an important role in all models. $K_{T}$ broadening 
can predict the shape of the data. Also taking $p$+Pb (shadow off)
as baseline in AMPT and HIJING highlights shadowing effect exclusively, other 
nuclear effects like multi-parton scattering phenomenon has considerable 
effects and can be viewed only with $p+p$ as baseline. 
To end with, further improvements are 
required in our parameter dependent models, to explain the experimental data properly. 
If results from high statistics data with improved 
uncertainty be available in future, we will improve these parameter dependent models to fit with data.


\end{document}